\documentclass[lettersize,journal]{IEEEtran}
\usepackage[font=footnotesize]{caption}
\usepackage{amsmath,amsfonts, mathtools}
\usepackage{dsfont, cases, tikz, subcaption}
\usepackage{hyperref}
\usepackage[capitalize,noabbrev,nameinlink]{cleveref}
\usepackage{algorithmic}
\usepackage[linesnumbered,ruled,vlined]{algorithm2e} % algorithm
\usepackage{url}
\usepackage{booktabs,multirow}
\usepackage[numbers,sort&compress]{natbib}
\usepackage{siunitx}

\crefformat{equation}{(#2#1#3)}
\Crefname{algocfline}{Algorithm}{Algorithms}
\Crefname{algocf}{line}{lines}
\Crefname{assumption}{Assumption}{Assumptions}
\crefrangeformat{assumption}{Assumptions~#3#1#4-#5#2#6}

\usepackage[acronyms, shortcuts]{glossaries}

\newcommand{\BibTeX}{\rm B\kern-.05em{\sc i\kern-.025em b}\kern-.08em\TeX}

\newtheorem{definition}{Definition}[section]

\newcommand{\ie}{\mathrm{i.e.}}
\newcommand{\st}{\mathrm{s.t.}~}

\newcommand{\ith}{~$i^\mathrm{th}$~}
\newcommand{\kth}{~$k^\mathrm{th}$~}

\newcommand{\argmin}{\mathop{\rm argmin}}

\newcommand{\R}{\mathbb{R}}
\newcommand{\X}{\mathcal{Z}}
\newcommand{\Lagrangian}{\mathcal{L}}

\newcommand{\equality}{g}
\newcommand{\equaldim}{m}
\newcommand{\GComp}{G}
\newcommand{\HComp}{H}
\newcommand{\inequality}{h}
\newcommand{\inequaldim}{p}

\newcommand{\bvar}{\mathbf{z}}
\newcommand{\tbvar}{\tilde{\mathbf{z}}}
\newcommand{\xstate}{x}
\newcommand{\control}{u}
\newcommand{\bstate}{\mathbf{x}}
\newcommand{\bcontrol}{\mathbf{u}}

\newcommand{\xdim}{n}

\newcommand{\horizon}{T}
\newcommand{\numplayers}{N}
\newcommand{\numlevels}{K}
\newcommand{\xposition}{p}
\newcommand{\velocity}{v}
\newcommand{\acceleration}{a}
\newcommand{\cost}[1]{J^{#1}}
\newcommand{\player}[1]{\text{R$#1$}}
\glsdisablehyper
\newacronym[longplural=open-loop Nash equilibria,plural=OLNE]{olne}{OLNE}{open-loop Nash equilibrium}
\newacronym[longplural=games of ordered preference,plural=GOOPs]{goop}{GOOP}{game of ordered preference}
\newacronym{micp}{MiCP}{mixed complementarity problem}
\newacronym{kkt}{KKT}{Karush-Kuhn-Tucker}
\newacronym{mpec}{MPEC}{mathematical program with equilibrium constraints}
\newacronym{mpcc}{MPCC}{mathematical program with complementarity constraints}
\newacronym{gnep}{GNEP}{generalized Nash equilibrium problem}
\newacronym[longplural=generalized Nash equilibria,plural=GNEs]{gne}{GNE}{generalized Nash equilibrium}
\newacronym{nlp}{NLP}{nonlinear programming}
\newacronym{mfcq}{MFCQ}{Mangasarian-Fromowitz constraint qualification}
\newacronym{licq}{LICQ}{linear independence constraint qualification}
\newacronym{cq}{CQ}{constraint qualification}

\begin{document}

\title{You Can't Always Get What You Want: \\ Games of Ordered Preference}

\author{Dong Ho Lee$^1$, Lasse Peters$^2$, and David Fridovich-Keil$^1$
        % <-this % stops a space
\thanks{Dong Ho Lee and David Fridovich-Keil are  with the Department of Aerospace Engineering and Engineering Mechanics, University of Texas at Austin, Austin 78712, USA (e-mail: {\tt leedh0124@utexas.edu; dfk@utexas.edu}).}
\thanks{Lasse Peters is with the Department of Cognitive Robotics (CoR), Delft
University of Technology, 2628 CD Delft, The Netherlands (e-mail: {\tt l.peters@tudelft.nl}).}
\thanks{This work was supported by a National Science Foundation CAREER award under Grant No. 2336840. (\it {Corresponding author: Dong Ho Lee.})}
% \thanks{This work has been submitted to the IEEE for possible publication. Copyright may be transferred without notice, after which this version may no longer be accessible.}% <-this % stops a space
}

% \author{Anonymous authors}



\maketitle

\begin{abstract}
We study noncooperative games, in which each player's objective is composed of a sequence of ordered---and potentially conflicting---preferences.
Problems of this type naturally model a wide variety of scenarios: for example, drivers at a busy intersection must balance the desire to make forward progress with the risk of collision.
Mathematically, these problems possess a nested structure, and to behave properly players must prioritize their most important preference, and only consider less important preferences to the extent that they do not compromise performance on more important ones.
We consider multi-agent, noncooperative variants of these problems, and seek generalized Nash equilibria in which each player's decision reflects both its hierarchy of preferences \emph{and} other players' actions.
We make two key contributions.
First, we develop a recursive approach for deriving the first-order optimality conditions of each player's nested problem.
Second, we propose a sequence of increasingly tight relaxations, each of which can be transcribed as a mixed complementarity problem and solved via existing methods.
Experimental results demonstrate that our approach reliably converges to equilibrium solutions that strictly reflect players' individual ordered preferences.
\end{abstract}

\begin{IEEEkeywords}
Non-cooperative games, lexicographic optimization, complementarity programming, multi-agent interaction
\end{IEEEkeywords}

%%%%%%%%%%%%%%%%%%%%%%%%%%%%%%%%%%%%%%%%%%%%%%%%%%%%%%%%%%%%%%%%%%%%%%%%
\section{Introduction} \label{sec:intro}
\begin{figure}[ht]
  \centering
  \includegraphics[width=1.0\linewidth]{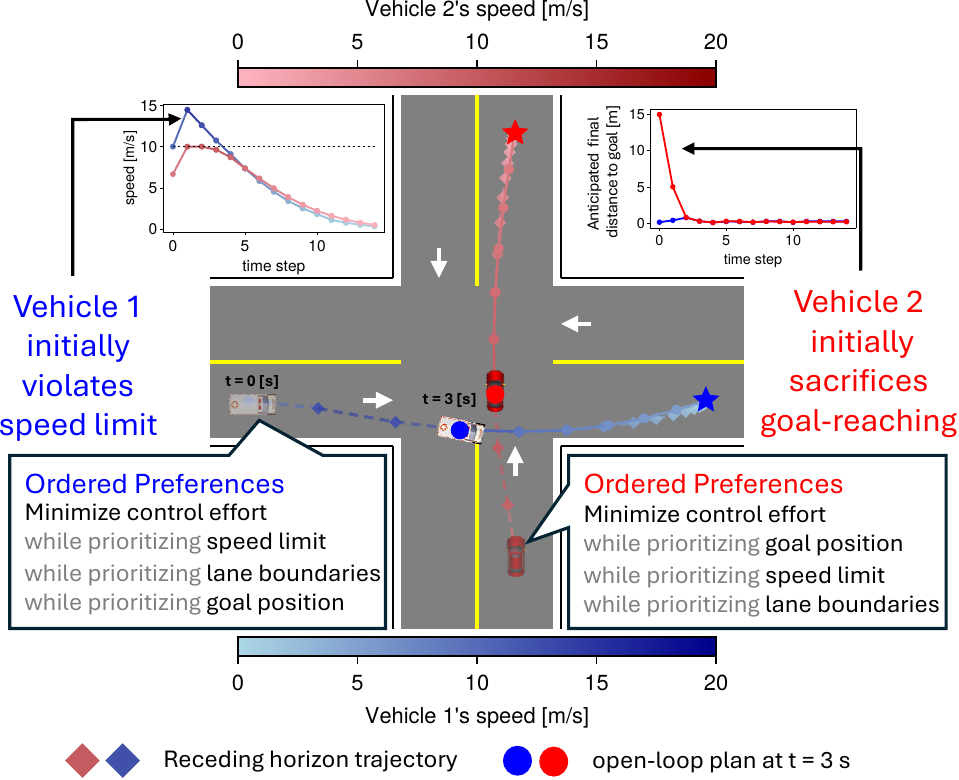}
  \caption{A two-vehicle intersection scenario involving four levels of preferences. The star indicates the goal position for each vehicle. Our \acf{goop} framework identifies equilibrium trajectories by selectively relaxing  less important preferences only when they compromise the performance of more important ones. In this figure, the dashed lines with diamond markers depict the complete closed-loop trajectories computed via receding horizon planning—where an open-loop plan is computed at each time step and only the first control action is implemented—while the solid lines with circle markers show a representative open-loop plan generated at $t = \SI{3}{\second}$. Additional subplots illustrate vehicle speeds and anticipated final distance to the goal in the closed-loop trajectories. Vehicle 1 (blue) initially violates the speed limit in order to satisfy goal-reaching at the final time step. On the other hand, Vehicle 2 (red) initially maintains their speed under the limit and sacrifices goal-reaching instead.}
  \label{fig:goop-intersection}
\end{figure}

\IEEEPARstart{I}{n} optimal decision-making, a user's preferences often reflect competing goals such as safety and efficiency.
For example, consider the intersection scenario in \cref{fig:goop-intersection} where each vehicle has a different order of preferences regarding reaching the goal, driving under the speed limit, driving within the lane, and minimizing fuel usage.
In such cases, treating all preferences as equally important can be problematic, especially when some preferences encode hard constraints, such as respecting lane boundaries. 
When formulated as an optimization problem, conflicting preferences can lead to infeasibility and ultimately cause solver failure.

In many cases---such as the autonomous driving example above---there is a clear hierarchy among the conflicting preferences.
A na\"ive approach to encode this concept of ordered preference is to construct a single objective function with weighted contributions from each preference, which can be adjusted manually or learned from data ~\cite{levine2012continuous}, ~\cite{ng2000algorithms}. However, such formulations can easily become ill-conditioned, and it is not always straightforward to design weights which yield the desired behavior.

Hierarchical optimization problems have been well studied in the operations research literature \cite{anandalingam1992hierarchical,lai1996hierarchical}. 
These problems are naturally characterized as a sequence of nested mathematical programs, where the decision variable at each level is constrained to be a minimizer of the problem at the level below.
Several studies such as \cite{allende2013solving,dempe2006optimality,dempe2013bilevel,dempe2015bilevel} have explored theoretical properties such as optimality conditions and constraint qualifications in bilevel settings.
Nested problems of this kind can also be solved via “lexicographic minimization,” in which each subproblem is addressed in order—from the lowest level to the highest level—while preserving the optimality of higher-priority preferences (at lower levels) by incorporating additional constraints \cite{kochenderfer2019algorithms, anilkumar2016lexicographic, khosravani2018application}.

While the lexicographic approaches produce solutions with desirable properties in single-agent settings, their computational methods do not readily extend to multi-agent domains. 
In single-agent contexts, hierarchical least-squares quadratic problems have been studied \cite{escande2014hierarchical}, particularly in the realm of real-time robot control. More generally, connectivity structures—often referred to as mathematical program networks—have also been characterized \cite{laine2024mathematical}.

While various methods to cope with hierarchical preferences have been developed---such as the aforementioned strategy of weighting agents' preferences according to their priority as in \cite{veer2023receding}---most focus on single-agent scenarios, and there are very limited results for multi-agent, noncooperative settings. 
For example, recent work  \cite{zanardi2021urban} applies lexicographic minimization to an urban driving game via an \emph{iterated best response} (IBR) scheme. 
However, this approach is limited to a certain class of games where IBR is guaranteed to converge. 
Follow-on work \cite{zanardi2021posetal} considers preferences which are only partially ordered, necessitating a substantially different solution approach.
Recent works \cite{Schwarting2019,fisac2019hierarchical} introduce social or game-theoretic models integrating high-level intent and compliance, while others \cite{wang2021game,geary2020resolving} study game-theoretic planning in competitive scenarios, but all assume scalarized or single-level objectives. In contrast, our work focuses on settings where agents’ decisions are governed by ordered preferences, allowing some to be selectively relaxed to achieve higher priority goals.

In this paper, we study multi-agent, game-theoretic variants of problems of (totally) ordered preference, which we refer to as \acp{goop}. Our contributions are twofold: \emph{(i)} We reformulate each agent's problem of ordered preference by sequentially replacing inner-level optimization problems with their corresponding \acl{kkt} conditions. 
This yields a \ac{mpcc} for each agent. 
\emph{(ii)} We develop a relaxation technique that smoothens the boundary of the feasible set in these problems in order to facilitate numerical computation.
From this set of relaxed \acp{mpcc}, we derive a single \acl{micp} whose solution is a (local) generalized Nash equilibrium solution of the original \ac{goop}.
We present experimental results which demonstrate that the proposed algorithm reliably converges to approximate generalized Nash solutions which reflect individual player's hierarchy of preferences, and compare the results with a family of penalty-based approximation baselines.

\section{Preliminaries and Related Work} \label{Preliminaries}
In this section, we introduce two important concepts underpinning our work and discuss the related literature in each area. In \cref{sec:mpcc}, we discuss how we formulate the problem of ordered preferences as a hierarchical optimization problem, and transcribe it as an \ac{mpcc}. Next, in \cref{sec:gne}, we introduce \acp{gnep} and discuss their relationship to \acp{micp}, for which an off-the-shelf solver is available.
% \begin{itemize}
%     \item Lexicographic Minimization 
%     \item Single agent case: hierarchical QP/lexicographic least squares
%     \item Multiagent case: Urban driving game/IBR
%     \item Mathematical Programming Network (MPN)

% \end{itemize}
\subsection{From Hierarchical Preferences to \acp{mpcc}}
\label{sec:mpcc}
We begin by discussing a \emph{single}-agent problem with \emph{two} levels; future sections will generalize to the $N$-agent, $K$-level setting.
% We begin with the simplest form of a hierarchical optimization problem, i.e., a problem with two levels.  
We use subscripts to denote the preference level and assume that a higher preference index indicates higher priority. In other words, the \emph{innermost} problem carries the highest level of preference.  
This yields a problem of the following form:
\begin{subequations}
\label{eqn:prelim-two_level}
\begin{align}
\min_{\bvar_1} \quad & \cost{}_1(\bvar_1) \\
\st \quad & \bvar_1 \in {\argmin_{\bvar_2}} \: \cost{}_2(\bvar_2) \label{eqn:prelim-inner_level_obj} \\
& \phantom{\st\quad} \st \bvar_2 \in \X, \label{eqn:prelim-inner_level_constraints}
\end{align}     
\end{subequations}
where $\forall i\in\{1,2\}, \bvar_i \in \R^\xdim, \cost{}_i(\cdot):\R^\xdim \to \R $, and the inner feasible set $\X$ is defined in terms of continuously differentiable functions $\equality:\R^\xdim \to \R^\equaldim$ and $\inequality:\R^\xdim \to \R^\inequaldim$ as $\X \coloneqq \{\bvar\in\R^\xdim~|~\equality(\bvar) = 0, \inequality(\bvar) \geq 0\}$. 
% \lasse{I think we shouldn't limit this discussion to 2 layers.}
This formulation captures the fact that any outer-level variables are constrained to be in the set of minimizers of the lower-level problem. 
By inspection, we can readily see that the inner problem is a constrained nonlinear program. In general, the \ac{kkt} conditions are only necessary for optimality, provided that some constraint qualifications are satisfied ~\cite{Pang07}.
If $\X$ is convex, then the \ac{kkt} conditions are also sufficient. 
% any local minimizer $\tilde{\bvar}$ of (\ref{eqn:prelim-inner_level_obj}) and (\ref{eqn:prelim-inner_level_constraints}) must satisfy the variational inequality defined by the set $\X$ and the mapping $\nabla \cost{}_2$, denoted as VI($\X, \nabla \cost{}_2$):
% \begin{equation}
%     (\mathbf{y} - \tilde{\bvar})^\top\nabla \cost{}_2 \geq 0, \quad \forall \mathbf{y} \in \X. 
% \end{equation}

% Because the inner problem defined by (\ref{eqn:prelim-inner_level_obj}) and (\ref{eqn:prelim-inner_level_constraints}) is a variational inequality when $\X$ is convex, the hierarchical problem is equivalent to a \ac{mpec}. 
% \lasse{Since there is no notion of other players yet, it doesn't make sense to talk about equilibria here. Either switch to bi-level optimization language or make equaiton (1) a game (latter probably better to save some space). If you stick to this two-level version, add a disclaimer that you limit it to two layers for notational simplicity}
The necessary conditions for optimality correspond to a \acf{micp}, which is the \ac{kkt} system comprised of primal ($\bvar_2$) and dual ($\lambda_2, \mu_2$) variables of the inner problem.
It is convenient to express the result in terms of the Lagrangian of the inner problem, defined as $\Lagrangian_2(\bvar_2, \lambda_2, \mu_2) \coloneqq \cost{}(\bvar_2) - \lambda_2^\top \inequality(\bvar_2) - \mu_2^\top \equality(\bvar_2)$: 
\begin{subequations}
\label{eqn:prelim-kkt-formulation}
\begin{align}
\min_{\bvar_1, \lambda_2, \mu_2} \quad & \cost{}_1(\bvar_1) \\
\st \quad & \nabla_{\bvar_2} \Lagrangian_2(\bvar_1, \lambda_2, \mu_2) = 0, \label{eqn:prelim-stationary} \\
&  0 \leq \inequality(\bvar_1) \ \bot \ \lambda_2 \geq 0, \label{eqn:prelim-complementarity}\\
&  \equality(\bvar_1) = 0. \label{eqn:prelim-equality}  
\end{align}
\end{subequations}
The optimization problem in (\ref{eqn:prelim-kkt-formulation}) is a single-level program that involves the Lagrange dual variables of the lower level problem in \cref{eqn:prelim-inner_level_obj} and \cref{eqn:prelim-inner_level_constraints} as primal variables. 
To be specific, the dual variables $(\lambda_2, \mu_2) \in \R^{\inequaldim+\equaldim}$, which are introduced at the inner problem, become primal variables $(\lambda_1, \mu_1)$ for the outer problem (in addition to $\bvar_1 \in \R^\xdim$).
We call these additional primal variables as the \emph{induced} primal variables since they are introduced in the process of building a single-level program. 
In particular, constraint \cref{eqn:prelim-stationary} refers to the stationarity condition of the Lagrangian function with respect to the primal variable ($\bvar_1$) of the inner level problem. 
Constraint \cref{eqn:prelim-complementarity} encodes the complementarity relationship between the inequality constraints in \cref{eqn:prelim-inner_level_constraints} and the associated dual variables. This constraint indicates that for each coordinate $i \in [\inequaldim]$, at least one of $\inequality_i(\bvar)$ and $\lambda_i$ is zero, while the other is nonnegative. 
Lastly, \cref{eqn:prelim-equality} is the equality constraint from \cref{eqn:prelim-inner_level_constraints}.

The reformulated problem in \cref{eqn:prelim-kkt-formulation}  is known as a \acf{mpcc}. In general, \acp{mpcc} are ill-posed as the complementarity constraints in \cref{eqn:prelim-complementarity} violate \acp{cq} such as the \ac{mfcq} and \ac{licq} at every feasible point ~\cite{scholtes2001convergence}. 
This inherent lack of regularity in the structure of \acp{mpcc} makes it difficult to use standard \ac{nlp} solvers directly. 
In particular, the absence of a \ac{cq} implies that the \ac{kkt} conditions of the reformulation in \cref{eqn:prelim-kkt-formulation} may no longer hold at a locally optimal solution. 
These theoretical and numerical difficulties led to the development of tailored theory and methods for solving \acp{mpcc} \cite{leyffer2006interior, schwartz2011mathematical, hoheisel2013theoretical, anitescu2005using, zemkohoo2020bilevel, nurkanovic2024solving}.
% Among various \ac{mpcc} methods, 
In this context, we develop a relaxation-based approach for solving \acp{goop}, which we explore in detail in \cref{sec:goop}.
% \david{conclude with a pointer saying that sometimes constraint qualifications break and this motivates us to introduce a relaxation method later on.}

% From here, talk about how this, i.e. approximation of MPEC becomes an MPCC, some features like treating duals as primals (MPCC-induced duals), if inner problem is nonconvex, then a solution for \ac{mpec} may be infeasible for the hierarchical problem
% General formulation in recursive form 
% We formally formulate a single-agent case as follows. 
% \begin{itemize}
%     \item Throughout the discussion, we follow the convention that the lowest level has the highest priority. 
%     \item Any sublevel problem is formulated so as to minimize the sum of slack variables $s_i$ where each slack variable determines the degree to which a particular priority constraint is to be violated. For completeness, additional constraints for slack nonnegativity are introduced as shown in... The slack variables for any sublevel are treated as additional primal variables.   
%     \item The optimization variables in any sublevel problem (including the outermost problem) is constrained to be in the set of minimizers of its lower-level problems. 
% \end{itemize}

\subsection{Generalized Nash Equilibrium Problems}
\label{sec:gne}
In this section, we formally introduce \acfp{gnep} and provide a brief overview of how local solutions may be identified %present a solution approach for a local \ac{gnep} point via first-order \ac{kkt} conditions ~
\cite{facchinei2010generalized}. 
A \ac{gnep} involves $\numplayers$ players, whose variables are denoted as $\bvar^i \in \R^{\xdim_i}$. 
The dimension of the game is $\xdim \coloneqq \sum_{i=1}^{N} \xdim_i$. 
We denote by  $\bvar^{\neg i} \in \R^{\xdim - \xdim_i}$ the state variables of all players except player \player{i}. 
Each player \player{i} has an objective function denoted by $\cost{i}(\bvar^i, \bvar^{\neg i})$ and a feasible set $\X^i(\bvar^{\neg i})$ on which their decisions depend.
Each feasible set is defined algebraically via (nonlinear) equality and/or inequality constraints~:~$\X^i(\bvar^{\neg i}) \coloneqq \{\bvar^i ~|~ \equality^i(\bvar^i, \bvar^{\neg i}) = 0, \inequality^i(\bvar^i, \bvar^{\neg i}) \geq 0\}.$ 
We call these constraints \emph{private} since they are ``owned'' by each player \player{i}.
Furthermore, we also consider constraints that are \emph{shared} among \numplayers~ players, which we denote as $\equality^s(\bvar) = 0, \inequality^s(\bvar) \geq 0$ where $\bvar \coloneqq [\bvar^1, \bvar^2, \dots, \bvar^\numplayers]^\top.$ 
For simplicity, we assume that these constraints are shared by \emph{all} players so that everyone is equally responsible for satisfying them.

\begin{definition}[Generalized Nash Equilibrium]
\label{def:gnep}
Mathematically, a \acf{gnep} is expressed via coupled optimization problems:
\begin{subequations}
\label{eqn:gnep}
\begin{alignat}{2}
    &\forall i \in [\numplayers] 
    &\quad &\left\{
    \label{eqn:gnep-private}
    \begin{aligned}
    &\min_{\bvar^i} && \cost{i}(\bvar^i, \bvar^{\neg i}) \quad  \\
    &~\textrm{s.t.} && \bvar^i \in \X^i(\bvar^{\neg i}) \quad 
    \end{aligned} \right. \\
     \label{eqn:gnep-shared-constraint}     
    &\textrm{s.t.} &\quad & \equality^s(\bvar) = 0, \quad \inequality^s(\bvar) \geq 0. \quad
\end{alignat}    
\end{subequations}

The \ac{gne} solution of \cref{eqn:gnep}, $\bvar^\ast \coloneqq [\bvar^{1\ast}, \dots, \bvar^{\numplayers^\ast}]^\top$, satisfies the inequality $\cost{i}(\bvar^i, \bvar^{\neg i\ast}) \geq \cost{i}(\bvar^\ast)$ for all feasible choices $\bvar^i \in \X^i(\bvar^{\neg i *})$, for all players $i \in [\numplayers]$.
This means that at equilibrium, no player has an incentive to unilaterally deviate from their equilibrium $\bvar^{i\ast}$. 
\end{definition}

In practice, it is intractable to solve for a (global) \ac{gne} solution. Instead, it is common to transcribe the formulation in \cref{eqn:gnep} as a \acf{micp} and use off-the-shelf solvers to find a \emph{local} \ac{gne} solution. In essence, solving this \ac{micp} is equivalent to finding a point that satisfies the system of first-order (\ac{kkt}) conditions of each player's optimization problem. 
In this paper, we use the PATH solver \cite{dirkse1995path}, which constructs an equivalent nonsmooth system of equations and solves them via a generalized Newton method.
We note that solving for \ac{gne} solutions via solving the corresponding \ac{micp} has been widely used in  \cite{rutherford2002mixed, cottle2009linear}.

% \lasse{todo: add more references that use this kind of approach (to support the claim that this is "common", micp methods)}

% implements a generalization of Newton's method with linesearch for an equivalent formulation of a complementarity problem as a nonsmooth system of equations.

\section{Games of Ordered Preferences} \label{sec:goop}

In this section, we formalize a variant of the hierarchical problems described in \cref{sec:mpcc} which extends to the multi-agent, noncooperative games of \cref{sec:gne}.
We term this multi-agent variant a \emph{\acl{goop}}. 

\subsection{Mathematical Formulation of \acp{goop}} 
We begin by introducing the mathematical formulation of \ac{goop} which we shall contextualize with a running example.

\subsubsection{General Formulation} Unlike the \ac{gnep} in \cref{def:gnep}, where each player's optimization problem is a standard \ac{nlp}, a \ac{goop} consists of $\numplayers$ optimization problems for each player,
but each player's problem is hierarchical, of the type discussed in \cref{sec:mpcc}. 
Each player's hierarchical problem may involve a different number of levels. 
To this end, we use $k^i \in [\numlevels^i]$ to denote the \kth level of preference for player \player{i}, where $\numlevels^i$ refers to the number of preferences for \player{i}.
Mathematically, we express a \ac{goop} as follows:
\begin{subequations}
\label{eqn:goop-K-level}
\begin{align}
    \label{eqn:goop-K-level-1}
    \quad \min_{\bvar^i_1} \quad & \cost{i}_1(\bvar^i_1, \bvar^{\neg i}_1) \\
    \label{eqn:goop-K-level-2}
    \st \quad & \bvar^i_1 \in \argmin_{\bvar^i_{2}} \: \cost{i}_{2}(\bvar^i_{2}, \bvar^{\neg i}_1) \\
    % \label{eqn:goop-K-level-3}
    & \phantom{\st\quad} \ddots \nonumber \\
    \label{eqn:goop-K-level-3}
    & \phantom{\st\quad} \st \bvar^i_{\numlevels^i - 1} \in \argmin_{\bvar^i_{\numlevels^i}} \: \cost{i}_{\numlevels^i}(\bvar^i_{\numlevels^i}, \bvar^{\neg i}_1) \\
    \label{eqn:goop-K-level-4}
    & \qquad\phantom{\st\quad\st\qquad} \st \bvar^i_{\numlevels^i} \in \X_{\numlevels^i}^i(\bvar^{\neg i}_1)\, \\
    \label{eqn:goop-shared-constraint}     
    & \equality^s(\bvar_1) = 0, \quad \inequality^s(\bvar_1) \geq 0. \quad
\end{align}
\end{subequations}

Here, $\X_{\numlevels^i}^i(\bvar^{\neg i}) \coloneqq \{\bvar^i\in\R^{\xdim_i}~|~\equality^i(\bvar^i, \bvar^{\neg i}) = 0, \inequality^i(\bvar^i, \bvar^{\neg i}) \geq 0\}$, and \cref{eqn:goop-shared-constraint} represents the shared constraints between \player{i} and the rest of the players.

\textbf{Running example.}
We will use the following 2-player running example to illustrate the \ac{goop} formalism.
We will study more complex interactions in \cref{sec:experiment}.

\begin{figure}[t]
  \centering
  \includegraphics[width=0.75\linewidth]{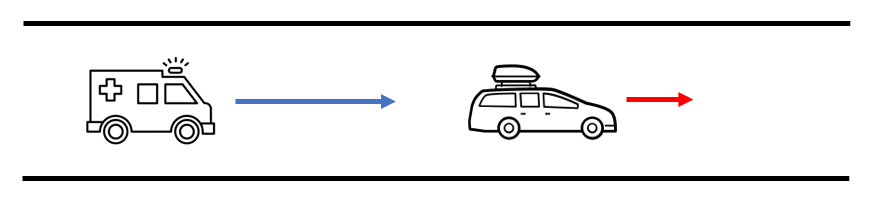}
  \caption{A highway driving scenario with 2 vehicles}
  \label{fig:highway}
\end{figure}

Consider the highway driving scenario of \cref{fig:highway}, in which $\numplayers=2$ vehicles must plan their future actions over the next $\horizon$ time steps.
In this example, vehicle $1$ is an ambulance and its highest priority preference is to reach a desired goal position.
Its secondary preference is to drive below the speed limit. 
In contrast, vehicle $2$ is a passenger car whose highest priority preference is to respect the speed limit, and whose secondary preference is to reach a goal location.
Both vehicles' lowest priority objective is to minimize their individual control effort, and no vehicle wants to collide.
These conflicting preferences make it natural to describe the interaction as a \ac{goop} \footnote{Although our running example considers only two levels of preferences, in practice one could also introduce another level which encodes an absolute maximum safety speed limit. This additional limit at a higher priority level would prevent the ambulance from exceeding the speed limit \emph{indefinitely}.}

We model each vehicle as a player in the game and denote the \ith vehicle's trajectory as $\bvar^i \coloneqq [\bstate^i, \bcontrol^i]^\top, \forall i \in [\numplayers]$.
Here, $\bstate^i \coloneqq [\xstate^i_1, \dots, \xstate^i_{\horizon}]^\top \in \R^{4\horizon}$ with $\xstate^i_t = [\xposition^i_{x, t}, \xposition^i_{y, t}, \velocity^i_{x, t}, \velocity^i_{y, t}]^\top \in \R^4$ encoding the state of \ith vehicle, comprised of position and velocity in the horizontal and vertical directions. 
% Without loss of generality, $\bstate^i$ is the state of an \ith vehicle aggregated over all time steps. 
Further, we denote a sequence of control inputs by $\bcontrol^i \coloneqq [\control^i_1, \dots, \control^i_{\horizon}]^\top \in \R^{2\horizon}$ where the \ith vehicle's control input at time $t$, $\control^i_t = [\acceleration_{x,t}^i, \acceleration_{y,t}^i]^\top \in \R^2$, is the acceleration in the horizontal and vertical directions, respectively.
Each vehicle follows double-integrator dynamics, discretized at a resolution $\Delta t$, i.e. 
\begin{equation}
\label{eqn:runexp-dynamics}
\xstate^i_{t+1} = \begin{bmatrix}
    1 & 0 & \Delta t & 0\\
    0 & 1 & 0 & \Delta t\\
    0 & 0 & 1 & 0\\
    0 & 0 & 0 & 1
\end{bmatrix} 
\underbrace{\begin{bmatrix}
    \xposition_{x,t}^i\\
    \xposition_{y,t}^i\\
    \velocity_{x,t}^i\\
    \velocity_{y,t}^i
\end{bmatrix}}_{\xstate_{t}^i} + \begin{bmatrix}
    \frac{1}{2}\Delta t^2 & 0\\
    0 & \frac{1}{2}\Delta t^2\\
    \Delta t & 0\\
    0 & \Delta t
\end{bmatrix} \underbrace{\begin{bmatrix}
    a^i_{x, t}\\
    a^i_{y, t}
\end{bmatrix}}_{u^i_t}\,.
\end{equation}
Note that \cref{eqn:runexp-dynamics} should be interpreted as equality constraints that partially define $\X_{\numlevels^i}^i(\bvar^{\neg i}_1)$ in \cref{eqn:goop-K-level-4}. 
Both vehicles must also drive within the highway lane in the horizontal direction. 
We encode this requirement as inequality constraints:
\begin{equation}
    \label{eqn:runexp-lane}
    \underline{\xposition}_y \leq \xposition_{y,t}^i \leq \overline{\xposition}_y
\end{equation}
The equality constraints \cref{eqn:runexp-dynamics} and inequality constraints \cref{eqn:runexp-lane} together specify the \emph{private} feasible set $\X_{\numlevels^i}^i(\bvar^{\neg i}_1)$ in \cref{eqn:goop-K-level-4}.
Both vehicles also share a collision-avoidance constraint:
\begin{equation}
\label{eqn:runexp-avoid-collision}
\inequality^s(\bstate^1, \bstate^2) = \left[\big( \xposition_{x,t}^1 - \xposition_{x,t}^2 \big)^2 + \big( \xposition_{y,t}^1 - \xposition_{y,t}^2 \big)^2 - d_{col}^2\right]_{t=1}^\horizon \in \R^\horizon
\end{equation}
where $d_{col}$ is the minimum distance between the two vehicles to avoid 
collision. 

To encode each player's individual ordered preferences, we define the following cost components:
\begin{subequations}
\label{eqn:runexp-priority-pref-1}
\begin{align}
    \cost{i}_\textrm{ctrl}(\bcontrol^i) &=  \sum_{t=1}^\horizon  \sum_{j \in \{x,y\}} \left(\acceleration^i_{j,t} \right)^2 \label{eqn:runexp-cost} \\
    \cost{i}_\mathrm{goal}(\bstate^i) &= \sum_{j \in \{x,y\}} [\hat\xposition^i_{j}-\xposition^i_{j,T}]_+ \label{eqn:runexp-reach-goal}  \\
     \cost{i}_\mathrm{obey}(\bstate^i) &= \sum_{\substack{t=1\\{j \in \{x,y\}}}}^\horizon [ \underline\velocity^i_{j}-\velocity^i_{j,t}]_+ + [\velocity^i_{j,t}-\overline\velocity^i_{j}]_+
     \label{eqn:runexp-speed-limit},
\end{align}
\end{subequations}
where $[\cdot]_+ := \max(0, \cdot)$, $(\xposition^i_{x,T}, \xposition^i_{y,T})$ represents the terminal position of the vehicle, $(\hat\xposition^i_{x}, \hat\xposition^i_{y})$ refers to the desired goal position, and $(\underline\velocity^i_{x}, \overline\velocity^i_{x})$ and $(\underline\velocity^i_{y}, \overline\velocity^i_{y})$ denote the lower and upper limits of the velocity in the horizontal and vertical directions.

Using these cost components, we define vehicle~1's ordered preferences to prioritize goal reaching \cref{eqn:runexp-reach-goal} over obeying the speed limit \cref{eqn:runexp-speed-limit}, $\ie$, $\cost{1}_2(\bstate^i) = \cost{1}_\mathrm{obey}(\bstate^1)$ and $\cost{1}_3(\bstate^1) = \cost{1}_\mathrm{goal}(\bstate^1)$.
In contrast, vehicle~2 prioritizes obeying the speed limit \cref{eqn:runexp-speed-limit} over goal reaching \cref{eqn:runexp-reach-goal}; i.e. $\cost{2}_2(\bstate^2) = \cost{2}_\mathrm{goal}(\bstate^2)$ and $\cost{2}_3(\bstate^2) = \cost{2}_\mathrm{obey}(\bstate^2)$.

Intuitively, the ambulance may violate the speed limit to reach the goal more quickly.
Similarly, the passenger car in front may pull to the side, de-prioritizing goal-reaching to yield to the ambulance, or temporarily violate the speed limit to avoid a collision.
Once the ambulance has passed, however, the car must strictly adhere to the speed limit.
\ac{goop} solutions naturally give rise to appropriate negotiation of preferences, relaxing less important preferences first when not all preferences can be perfectly satisfied.   

\begin{figure}[t]
  \centering
  \includegraphics[width=0.7\linewidth]{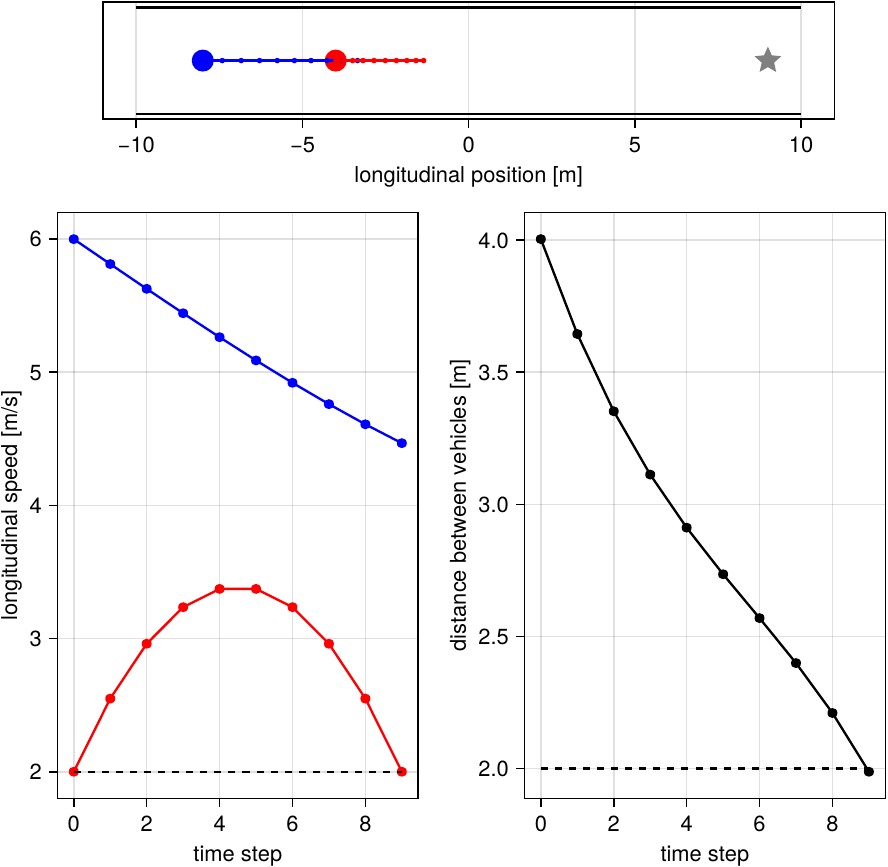}
  \caption{A \ac{goop} solution of the running example. The grey marker is the goal position for both vehicles. The dashed line in the left plot (speed) indicates the maximum speed limit. The dashed line in the right plot (distance between the vehicles) represents the minimum safe distance for collision avoidance.}
  \label{fig:sample-openloop}
\end{figure}

To support this intuition, we provide a sample solution for the highway running example. 
\cref{fig:sample-openloop} shows the interaction between two vehicles with different priorities for this scenario where we consider horizontal dynamics only.  
Vehicle 1 (blue) prioritizes minimizing the distance to the goal at the final time step. However, it slows down in order to avoid collision with vehicle 2. 
Vehicle 2 (red) prioritizes driving within the maximum speed limit, but to avoid collision with the fast-approaching Vehicle 1 (blue), it temporarily exceeds this limit. 
%The distance between the two gradually decreases until, at the final time step, both vehicles are at the minimum safe distance. 
Here, \ac{goop} allows optimal violations of preferences to satisfy hard constraints like collision avoidance. 

%\begin{remark}
%\label{remark1}
%Note that the presence of the $\max(\cdot)$ in %\cref{eqn:runexp-reach-goal,eqn:runexp-speed-limit} makes these objectives non-smooth. 
%\Cref{sec:experiment} describes a slack variable %reformulation which renders them smooth.   
%\end{remark}

\subsubsection{From hierarchical to single-level}
Next, we discuss how to derive first-order necessary conditions for \ac{goop}.
We shall use these conditions to identify equilibrium solutions in \cref{sec:goop-mpcc}.

Following the procedure in \cref{sec:mpcc}, we may transcribe \player{i}'s hierarchical problem \cref{eqn:goop-K-level} into a single level.
To do this, we will successively replace each nested problem within \cref{eqn:goop-K-level-2,eqn:goop-K-level-3,eqn:goop-K-level-4} with its corresponding \ac{kkt} conditions, starting from the inner-most problem, which encodes the highest priority preference.
As a result of this operation, we obtain a \acf{mpcc} of the following form:
\begin{subequations}
\label{eqn:goop-MPCC}
\begin{align}
\tbvar^{i*}_1 \in \arg\min_{\tbvar^i_1} & \ \cost{i}_1(\tbvar^i_1, \tbvar^{\neg i*}_1) \\
\st & \ \equality^i(\tbvar^i_1, {\tbvar^{\neg i*}_1}) = 0, ~\inequality^i(\tbvar^i_1, \tbvar^{\neg i*}_1) \geq 0, \label{eqn:goop-equality-inequality}\\
    & \GComp^i(\tbvar^i_1, \tbvar^{\neg i*}_1) \geq 0, ~\HComp^i(\tbvar^i_1, \tbvar^{\neg i*}_1) \geq 0, \label{eqn:goop-MPCC-nonnegativity} \\
    & \GComp^i(\tbvar^i_1, \tbvar^{\neg i*}_1)^\top \HComp^i(\tbvar^i_1, \tbvar^{\neg i*}_1) = 0, \label{eqn:goop-MPCC-complementarity} \\
    & \equality^s(\tbvar^i_1, \tbvar^{\neg i*}_1) = 0, \ \inequality^s(\tbvar^i_1, \tbvar^{\neg i*}_1) \geq 0.     \label{eqn:goop-MPCC-shared-constraint}   
\end{align}
\end{subequations}
and we interpret the constraints in \cref{eqn:goop-equality-inequality,eqn:goop-MPCC-nonnegativity,eqn:goop-MPCC-complementarity} as a specification of \player{i}'s private constraint set $\X^i(\tbvar^{\neg i}_1)$ for the \ac{gnep} in \cref{eqn:gnep-private}, and \cref{eqn:goop-MPCC-shared-constraint} encode the shared constraints in \cref{eqn:gnep-shared-constraint}. 

Note that problem \cref{eqn:goop-MPCC} involves new variables $\tbvar^i_1,~\forall i \in [\numplayers]$.
These include the original primal variables $\bvar^i_1$ along with additional variables---the dual variables from lower-level problems in \cref{eqn:goop-K-level}---induced by the aforementioned recursive procedure. In particular, $\tbvar^i_1 \coloneqq \left[\bvar^i_1,\lambda^i_2, \mu^i_2, \dots, \lambda^i_{K^i}, \mu^i_{K^i} \right]^\top, \forall i \in [\numplayers]$, and the variables ($\lambda^i_2, \mu^i_2, \dots, \lambda^i_{K^i}, \mu^i_{K^i}$) are Lagrange multipliers from the \ac{kkt} conditions of lower-level problems.
The functions $\equality^i(\tbvar^i_1, \tbvar^{\neg i}_1), \inequality^i(\tbvar^i_1, \tbvar^{\neg i}_1), \GComp^i(\tbvar^i_1, \tbvar^{\neg i}_1)$ and $\HComp^i(\tbvar^i_1, \tbvar^{\neg i}_1)$ collect equality and inequality constraints that arise throughout.
% Each player's Lagrangian incorporates their private and shared constraints in \cref{eqn:goop-MPCC} as:
% \begin{equation}
% \label{eqn:goop-MPCC-Lagrangian}
% \begin{split}
%     &\Lagrangian^i(\tbvar^i_1, \lambda^i_1, \mu^i_1, \lambda^s, \mu^s) =  \, \cost{i}_1(\tbvar^i_1, \tbvar^{\neg i}_1) -  \begin{bmatrix}
%         \inequality^i(\tbvar^i_1, \tbvar^{\neg i}_1) \\
%         \GComp^i(\tbvar^i_1, \tbvar^{\neg i}_1) \\
%         \HComp^i(\tbvar^i_1, \tbvar^{\neg i}_1)
%     \end{bmatrix}^\top \lambda^{i}_1 \\
%      &-\ \begin{bmatrix}
%         \equality^i(\tbvar^i_1, \tbvar^{\neg i}_1) \\
%         \GComp^{i}(\tbvar^i_1, \tbvar^{\neg i}_1)^\top \HComp^i(\tbvar^i_1, \tbvar^{\neg i}_1) 
%     \end{bmatrix}^\top\mu^{i}_1 - \inequality^s(\tbvar_1)^\top \lambda^s - \equality^s(\tbvar_1)^\top \mu^s.
% \end{split}
% \end{equation}
% Because the constraints in \cref{eqn:goop-MPCC-shared-constraint} are shared among all players, the players share the same Lagrange multiplier ($\lambda^s,\mu^s$) for each constraint. 
% As we shall see in \cref{sec:from mpcc to micp}, the Lagrangian functions for each player in \cref{eqn:goop-MPCC-Lagrangian} plays an important role in solving for \ac{gne} solutions.

For clarity, we present an explicit formulation of \cref{eqn:goop-MPCC} in the following running example.
% We will discuss the encoding of shared constraints in \cref{sec:shared constraints in MPCC}.

\textbf{Running example.} For our running example, we have three priority levels for each vehicle, $\ie, \numlevels^i = 3, ~\forall i \in [2]$. For this simple case, we can see how the dual variables become \emph{induced} primal variables for the outermost problem. Beginning with the innermost level ($k^i=3$), the intermediate level ($k^i=2$) problem becomes: 
\begin{subequations}
\label{eqn:goop-intermediate-level}
\begin{align}
\min_{\bvar^i_2, {\color{blue}{\lambda^i_3}}, {\color{blue}{\mu^i_3}}} \quad & \cost{i}_2(\bvar^i_2, \tbvar^{\neg i}_1) \\
\st \quad & \nabla_{\bvar^i_3} \Lagrangian^i_3(\bvar^i_2, \tbvar^{\neg i}_1, {\color{blue}{\lambda^i_3}}, {\color{blue}{\mu^i_3}}) = 0, \label{eqn:goop-intermediate-level-stationary} \\
&  0 \leq \inequality^i(\bvar^i_2, \tbvar^{\neg i}_1) \ \bot \ {\color{blue}{\lambda^i_3}} \geq 0, \label{eqn:goop-intermediate-level-complementarity}\\
&  \equality^i(\bvar^i_2, \tbvar^{\neg i}_1) = 0. \label{eqn:goop-equality}  
\end{align}
\end{subequations}
The \ac{kkt} conditions for \cref{eqn:goop-intermediate-level} define the feasible set of the outermost ($k^i=1$) problem: 
\begin{subequations}
\label{eqn:goop-final-level}
\begin{align}
\min_{\bvar^i_1, {\color{blue}{\lambda^i_3}}, {\color{blue}{\mu^i_3}}, {\color{red}{\lambda^i_2}}, {\color{red}{\mu^i_2}}} & \ \cost{i}_1(\bvar^i_1, \tbvar^{\neg i}_1) \\
\st \quad & \nabla_{\bvar^i_2, {\color{blue}{\lambda^i_3}}, {\color{blue}{\mu^i_3}}} \Lagrangian^i_2(\bvar^i_1, \tbvar^{\neg i}_1, {\color{blue}{\lambda^i_3}}, {\color{blue}{\mu^i_3}}, {\color{red}{\lambda^i_2}}, {\color{red}{\mu^i_2}}) = 0, \label{eqn:goop-intermediate-level-stationary-1} \\
 & \nabla_{\bvar^i_3} \Lagrangian^i_3(\bvar^i_1, \tbvar^{\neg i}_1, {\color{blue}{\lambda^i_3}}, {\color{blue}{\mu^i_3}}) = 0, \label{eqn:goop-intermediate-level-stationary-2} \\
&  0 \leq \inequality^i(\bvar^i_1, \tbvar^{\neg i}_1) \ \bot \ {\color{red}{\lambda^{i,1}_2}} \geq 0 \label{eqn:goop-intermediate-level-complementarity-1}, \\ 
&  0 \leq {\color{blue}{\lambda^i_3}} \ \bot \ {\color{red}{\lambda^{i,2}_2}} \geq 0 \label{eqn:goop-intermediate-level-complementarity-2}, \\
& \inequality^i(\bvar^i_1, \tbvar^{\neg i}_1)^\top {\color{blue}{\lambda^i_3}} = 0, ~\equality^i(\bvar^i_1, \tbvar^{\neg i}_1) = 0, \label{eqn:goop-equality-1}\\
% & {\color{red}{\mu^{i,1}_2}} \cdot \inequality^i(\bvar^i_1, \tbvar^{\neg i}_1)^\top {\color{blue}{\lambda^i_3}} = 0, \label{eqn:goop-equality-2}\\
% & \equality^i(\bvar^i_1, \tbvar^{\neg i}_1)^\top {\color{red}{\mu^{i,2}_2}} = 0 \label{eqn:goop-equality-3} \\
& \equality^s(\tbvar_1) = 0, \quad \inequality^s(\tbvar_1) \geq 0 \label{eqn:goop-shared-constraint-running-example}.
\end{align}
\end{subequations}
where ${\color{red}{\lambda^i_2}} = \left[{\color{red}{\lambda^{i,1}_2}},{\color{red}{\lambda^{i,2}_2}}\right]^\top$, ${\color{red}{\mu^i_2}} = \left[{\color{red}{\mu^{i,1}_2}},{\color{red}{\mu^{i,2}_2}}, {\color{red}{\mu^{i,3}_2}}\right]^\top$ denote dual variables for the inequality and equality constraints (respectively) of the intermediate level problem. Note that the Lagrangian for the outermost problem is 
$\Lagrangian^i_2(\bvar^i_2, \tbvar^{\neg i}_1, {\color{blue}{\lambda^i_3}}, {\color{blue}{\mu^i_3}}, {\color{red}{\lambda^i_2}}, {\color{red}{\mu^i_2}}) = \cost{i}_2(\bvar^i_2, \tbvar^{\neg i}_1) - \inequality^i(\bvar^i_2, \tbvar^{\neg i}_1)^\top {\color{red}{\lambda^{i,1}_2}} - {\color{blue}{\lambda^i_3}}^\top {\color{red}{\lambda^{i,2}_2}} - \nabla_{\bvar^i_3} \Lagrangian^i_3(\bvar^i_2, \tbvar^{\neg i}_1, {\color{blue}{\lambda^i_3}},{\color{blue}{\mu^i_3}})^\top {\color{red}{\mu^{i,1}_2}} - \equality^i(\bvar^i_2, \tbvar^{\neg i}_1)^\top {\color{red}{\mu^{i,2}_2}} - \inequality^i(\bvar^i_2, \tbvar^{\neg i}_1)^\top {\color{blue}{\lambda^i_3}} \cdot {\color{red}{\mu^{i,3}_2}}$. 
Observe that the formulation in \cref{eqn:goop-final-level} is in the form of an \ac{mpcc} as given in \cref{eqn:goop-MPCC}.
To be specific, we have that $\equality^i(\tbvar^i_1, \tbvar^{\neg i}_1)$ consists of the equality constraints \cref{eqn:goop-intermediate-level-stationary-1,eqn:goop-intermediate-level-stationary-2,eqn:goop-equality-1}.
The shared constraints in \cref{eqn:goop-MPCC-shared-constraint} are identical to \cref{eqn:goop-shared-constraint-running-example} and
the complementarity constraints in \cref{eqn:goop-MPCC} correspond to:
\begin{equation}
     \GComp^i(\tbvar^i_1, \tbvar^{\neg i}_1) = \begin{bmatrix}
        \inequality^i(\bvar^i_1, \tbvar^{\neg i}_1) \\
        \color{blue}{\lambda^i_3}
     \end{bmatrix}, \
     \HComp^i(\tbvar^i_1, \tbvar^{\neg i}_1) = \begin{bmatrix}
        \color{red}{\lambda^{i,1}_2} \\
        \color{red}{\lambda^{i,2}_2}
     \end{bmatrix}.
\end{equation}
Next, we discuss how to solve problems of the form~\cref{eqn:goop-MPCC} numerically.
%\lasse{note to self: continue here}
\subsection{Numerical Solution of \ac{goop}}
\label{sec:goop-mpcc}
% In this section, we propose a relaxation algorithm which mitigates the aforemen \acp{goop}. 
\subsubsection{MPCC Relaxation}
As noted earlier in \cref{sec:mpcc}, the \ac{mpcc} in \cref{eqn:goop-MPCC} can be numerically challenging to solve due to irregularities in the geometry of the feasible set.
Therefore, we propose a relaxation scheme that mitigates the aforementioned issue by solving a sequence of \acp{goop} which are regularized by altering the complementarity constraints in \cref{eqn:goop-MPCC-complementarity}. 
To this end, we replace the equality constraint in \cref{eqn:goop-MPCC-complementarity} with an inequality as follows:
%\begin{equation}
%     G^i(\tbvar^i_1, \tbvar^{\neg i}_1)^\top H^i(\tbvar^i_1, \tbvar^{\neg i}_1) \leq 0. \label{eqn:goop-MPCC-complementarity-equivalent}
%\end{equation}
%With $\sigma > 0$, we can further replace \cref{eqn:goop-MPCC-complementarity-equivalent} with 
\begin{equation}
     G^i(\tbvar^i_1, \tbvar^{\neg i}_1)^\top H^i(\tbvar^i_1, \tbvar^{\neg i}_1) \leq \sigma. \label{eqn:goop-MPCC-complementarity-relaxed}
\end{equation}
When $\sigma=0$, this constraint encodes the original complementarity condition.
For $\sigma>0$, this reformulation enlarges the feasible set and ensures that it has a nonempty interior. 
% as shown in \cref{fig:enlarged-complementarity-set}.

% \begin{figure}[t]
% \centering
% \begin{tikzpicture}
%     % Define the range of x-values and y-values
%     \def\xmin{1/\ymax} % Corresponding x value when y = ymax
%     \def\xmax{2.5}
%     \def\ymax{2.5}
%     \def\xrect{1/2.5+0.01} % Define the end of the rectangle in x

%     % Draw and shade the area under the curve y = 1/x, stopping at ymax
%     \fill[orange!50] plot[domain=\xmin:\xmax, smooth] (\x, {1/\x}) -- (\xmax, 0) -- (\xmin, 0) -- cycle;
%     \fill[orange!50] (0, 0) rectangle (\xrect, \ymax);

%     % Draw the curve y = 1/x, stopping at ymax
%     \draw[thick, orange] plot[domain=\xmin:\xmax, smooth] (\x, {1/\x});

%     % Axes
%     \draw[->] (-0.1, 0) -- (\xmax+0.1, 0) node[right] {$G^i(x)$}; % x-axis with label
%     \draw[->] (0, -0.1) -- (0, \ymax+0.1) node[above] {$H^i(x)$}; % y-axis with label

%     % Label for the curve
%     \node at (2.2, 1.5) {$G^i(x)^\top H^i(x) \leq \sigma$};

% \end{tikzpicture}
% \caption{Enlarged feasible set from relaxed complementarity conditions.}
% \label{fig:enlarged-complementarity-set}
% \end{figure}

Using this relaxation scheme, the \ac{mpcc} in \cref{eqn:goop-MPCC} becomes
\begin{subequations}
\label{eqn:goop-relaxed-MPCC}
\begin{align}
\tbvar^{i*}_1 \in \arg\min_{\tbvar^i_1} & \ \cost{i}_1(\tbvar^i_1, \tbvar^{\neg i*}_1) \\
\st & \ \equality^i(\tbvar^i_1, \tbvar^{\neg i*}_1) = 0, ~\inequality^i(\tbvar^i_1, \tbvar^{\neg i*}_1) \geq 0, \label{eqn:goop-relaxed-MPCC-equality-inequality}\\
    & \GComp^i(\tbvar^i_1, \tbvar^{\neg i*}_1) \geq 0, ~\HComp^i(\tbvar^i_1, \tbvar^{\neg i*}_1) \geq 0, \label{eqn:goop-relaxed-MPCC-nonnegativity} \\
    & \color{blue}{\GComp^i(\tbvar^i_1, \tbvar^{\neg i*}_1)^\top \HComp^i(\tbvar^i_1, \tbvar^{\neg i*}_1) \leq \sigma}, \label{eqn:goop-relaxed-MPCC-complementarity} \\
    & \equality^s(\tbvar_1, \tbvar^{\neg i*}_1) = 0, \ \inequality^s(\tbvar_1, \tbvar^{\neg i*}_1) \geq 0.     \label{eqn:goop-relaxed-MPCC-shared-constraint}     
\end{align}
\end{subequations}

%The resulting single-level game comprises of each player \player{i}'s optimization problem in the form of \cref{eqn:goop-relaxed-MPCC}.
\subsubsection{From relaxed \ac{mpcc} to \ac{micp}}
\label{sec:from mpcc to micp}
To solve this transcribed game, we formulate the \ac{kkt} conditions of the coupled optimization problem, i.e.,
\begin{equation}\label{eqn:goop-micp}
\begin{bmatrix}
    \nabla_{\tbvar^1_1}\tilde\Lagrangian^1 \\
    \equality^1 \\
    \vdots \\
    \nabla_{\tbvar^\numplayers_1}\tilde\Lagrangian^\numplayers \\
    \equality^\numplayers \\
    \equality^s
\end{bmatrix} = 0~\mathrm{and}~
0 \leq \begin{bmatrix}
    c^1 \\
    \vdots \\
    c^\numplayers \\
    \inequality^s
\end{bmatrix}  \bot \ \boldsymbol{\lambda} \geq 0, 
\end{equation}
where
\begin{equation}
\label{eqn:goop-Relaxed-MPCC-Lagrangian}
\begin{split}
    \tilde\Lagrangian^i(\tbvar^i_1, \tilde\lambda^i_1, \tilde\mu^i_1, \lambda^s, \mu^s) =  \, \cost{i}_1(\tbvar^i_1, \tbvar^{\neg i}_1) -  c^i(\tbvar^i_1, \tbvar^{\neg i}_1)^\top \tilde\lambda^{i}_1 \\
     -\equality^i(\tbvar^i_1, \tbvar^{\neg i}_1) ^\top\tilde\mu^{i}_1 - \inequality^s(\tbvar_1)^\top \tilde\lambda^s - \equality^s(\tbvar_1)^\top \tilde\mu^s,
\end{split}
\end{equation}
is the Lagrangian of the $i$th player's problem with Lagrange multipliers ($\tilde \lambda^i_1, \tilde \mu^i_1, \tilde \mu^s, \tilde \lambda^s$),
\begin{equation}
c^i(\tbvar^i_1, \tbvar^{\neg i}_1) \coloneqq  
\begin{bmatrix}
    \inequality^i(\tbvar^i_1, \tbvar^{\neg i}_1) \\
    G^i(\tbvar^i_1, \tbvar^{\neg i}_1) \\
    H^i(\tbvar^i_1, \tbvar^{\neg i}_1) \\
    \sigma - G^i(\tbvar^i_1, \tbvar^{\neg i}_1)^\top H^i(\tbvar^i_1, \tbvar^{\neg i}_1)
\end{bmatrix}
\end{equation}
denotes the aggregated vector of player $i$th (private) inequality constraints in \cref{eqn:goop-relaxed-MPCC-equality-inequality,eqn:goop-relaxed-MPCC-nonnegativity,eqn:goop-relaxed-MPCC-complementarity},
and $\boldsymbol{\lambda}$ denotes the aggregation of all players Lagrange multiplies associated with inequality constraints.
%In order to view the relaxed problem as a standard \acf{micp}---for which there are existing solvers---we introduce new variables $\mathbf{q} \coloneqq [\tbvar, \boldsymbol{\mu}]^\top$ and $\boldsymbol{\lambda}$, where $\boldsymbol{\mu}$ and $\boldsymbol{\lambda}$ are the aggregation of all the Lagrange multipliers associated with each player \player{i}'s equality and inequality constraints (including shared constraints), respectively. The \ac{kkt} system of the relaxed $\numplayers$-player \ac{goop} can then be written as: 
%\begin{equation}
%\begin{bmatrix}
%    \nabla_{\tbvar^1_1}\tilde\Lagrangian^1 \\
%    \equality^1 \\
%    \vdots \\
%    %\nabla_{\tbvar^\numplayers_1}\tilde\Lagrangian^\numplayers \\
%    \equality^\numplayers \\
%    \equality^s
%\end{bmatrix} = 0~\mathrm{and}~
%0 \leq \begin{bmatrix}
%    c^1 \\
%    \vdots \\
%    c^\numplayers \\
%    \inequality^s
%\end{bmatrix}  \bot \ \boldsymbol{\lambda} \geq 0, 
%\end{equation}
%which is the standard \ac{micp} in \cite[Definition 1.1.6]{Pang07}. 
The resulting \ac{kkt} conditions in \cref{eqn:goop-micp} take the form of a standard \ac{micp} \cite[Definition 1.1.6]{Pang07} for which off-the-shelf solvers exist.

\subsubsection{Proposed Algorithm}
% In \cref{eqn:goop-MPCC-complementarity-relaxed}, $\sigma > 0$ is the relaxation factor and is initialized as a small positive number.
% We apply the relaxation factor at the complementarity constraints throughout all (inner and outermost) levels and treat each relaxed complementarity constraint as an inequality constraint as we follow the procedure in \cref{sec:mpcc}. \david{bit complicated - reassess after changes above}
% We denote the resulting \ac{goop} as \ac{goop}($\sigma$).

%\lasse{needs to be said somewhere but not earlier than here: We denote the resulting \ac{goop} as \ac{goop}($\sigma$).}
With the above relaxation scheme at hand, we numerically solve the original (unrelaxed) \ac{goop} via a sequence of successively tightened relaxations~\cref{eqn:goop-micp}; i.e. with $\sigma$ successively approaching zero.

Our proposed procedure is summarized in \cref{algorithm} in which \ac{goop}($\sigma$) denotes the relaxed \ac{micp} at tightness $\sigma$.
Specifically, we start by initializing $\tbvar$ as a vector of zeros of the appropriate dimension and setting $\sigma$ as a small positive number.
We then solve the resulting \ac{micp} using the PATH solver \cite{dirkse1995path}, and repeat for successively smaller $\sigma$ using each solution $\tbvar$ as an initial guess for the next round. 
In this way, we gradually drive $\sigma$ to zero and find a local \ac{gne} solution such that the maximum violation of complementarity, $\max_j \{ G^i_j(\tbvar^i_1, \tbvar^{\neg i}_1)H^i_j(\tbvar^i_1, \tbvar^{\neg i}_1) \}_{i=1}^{N}$ is below a certain tolerance, $\gamma > 0$.
The convergence of such annealing procedures has been widely studied in the context of general \acp{mpec} and \acp{mpcc}. 
Under tailored constraint qualifications outlined in \cite{scholtes2001convergence,hoheisel2013theoretical}, the stationary points of the relaxed problems converge to a weak stationary point of the underlying \ac{mpec}. 
For more details on convergence results, we refer readers to \cite{scholtes2001convergence,hoheisel2013theoretical}.
If, at any iteration, the current solution does not change significantly from the previous one, $\ie$, by more than a fixed tolerance $\epsilon > 0$, we consider the solution has converged.
We refer to \cite{scholtes2001convergence} for guidelines on selecting ($\sigma, \kappa, \gamma, \epsilon$) in \cref{algorithm}.
%We guide setting the values of these hyper-parameters in \cref{sec:experiment}.\lasse{do we?}

\begin{algorithm}
\label{algorithm}
    \caption{Relaxed Game of Ordered Preferences}
        $\bvar_{0}, \sigma_{0}, \kappa, \gamma, \epsilon \leftarrow$ 
        initial guess, relaxation factor, update factor, complementarity tolerance, converged tolerance \\
    Set $k \gets 1$ \\
    \While{$\max_j \{ G^i_j(\bvar_k)H^i_j(\bvar_k) \}_{i=1}^{N} \geq \gamma $ or $k = 1$}{
        $\bvar_{k} \leftarrow \text{solution of \ac{goop}($\sigma_{k-1}$) initialized at}~\bvar_{k-1}$ \\
        % $\bvar_{k} \leftarrow \bvar_{k+1}$  \tcp*{Solve relaxed problem \ac{goop}($\sigma_k$)} 
        \If{$\max_j \{ G^i_j(\bvar_k)H^i_j(\bvar_k) \}_{i=1}^{N} \leq \gamma $}{
            \bf{break} \tcp*{Solution is found} 
        }
        \ElseIf{ $||\bvar_{k} - \bvar_{k-1}||_2 < \epsilon$}{
            \bf{break}  \tcp*{Low precision solution} 
        }
        \Else{
            $\sigma_{k} \leftarrow \kappa\sigma_{k-1}$ \tcp*{Reduce $\sigma \downarrow 0$} 
            $k \leftarrow k+1$ 
        }
    }
    \Return: $\bvar_{k}, \sigma_k, \max_j \{ G^i_j(\bvar_k)H^i_j(\bvar_k) \}_{i=1}^{N}$
\end{algorithm}

% \begin{algorithm}
%     \caption{Relaxed Receding Speculator Game}
%     \label{alg1}
%     $S_{0}, C_{0}, \alpha_{0}, \widehat{p}^{S}_{0}, \widehat{p}^{C}_{0}, \tol , T \leftarrow$
%         Initial State, Price Forecasts, Solver Tolerance, Time Horizon \\
%     \While{$t < \infty$}{
%       $\beps \leftarrow \bm{1}$  \tcp*{Reset $\beps$ to Ones Vector}\\
%       \For{ \( n = 1 \) \textbf{ to } \( 10 \)}{
%         $\zstar = \begin{pmatrix} \bm{x^{*}}, \bm{y^{*}} \end{pmatrix} \leftarrow$ Solve relaxed Problem \cref{relaxed} \\
%         \If{ $||\zstar - \zlast||_2 < \tol$}{
%                 $\alpha \leftarrow$  Update rate from first time step of \( x^{*} \) \\
%                  \bf{break}
%             }
%             \Else{
%                 $\beps \leftarrow \frac{1}{2^{n}}\beps$ \tcp*{Gradually reduce $\beps \rightarrow 0$} \\
%                 $\zlast \leftarrow \zstar$
%             }
%       }
%       $\Delta, p^{C}_{t}, p^{S}_{t} \leftarrow$ Get Market Observations \\
%       $\widehat{p}^{C}_{t}, \widehat{p}^{S}_{t} \leftarrow $ Update forecasts from $p^{C}_{t}, p^{S}_{t}$ \\
%       $S_{t}, C_{t} \leftarrow$ From $p^{C}_{t}, p^{S}_{t}, \Delta , \alpha$ using \cref{supply}, \cref{collateral}
%     }
% \end{algorithm}

\section{Experiments} \label{sec:experiment}
This section evaluates the performance of the proposed \ac{goop} approach in a Monte Carlo study and compares it with a baseline that encodes the ordered preferences via penalty-based scalarization in a non-hierarchical game formulation.
These experiments are designed to support the claims that (\romannumeral 1) \ac{goop} reliably reflect agents' individual ordered preferences and that (\romannumeral 2) penalty-based approximate scalarization schemes fail to capture such solutions.
% \dongho{To this end, we (i) compare the numerical values of agents' preferences at each level, and (ii) compare the equilibrium values of primal variables found by each method.}
Finally, we also present a scenario with more complex dynamics and preferences to illustrate the practical generality of the \ac{goop} framework.

\subsection{Experiment Setup}
% \lasse{describe experiment scenario first}
\textbf{Evaluation Scenario.}
Our experiment extends the previous running example of highway driving scenario, where we consider $\numplayers=3$ vehicles:  vehicle 1 (blue) is an ambulance that wishes to travel at high speed, and vehicles 2 (red) and 3 (green) are passenger cars just ahead of the ambulance. 
Each vehicle adheres to a specific hierarchy of preferences, as outlined in \cref{eqn:runexp-priority-pref-1}.
Note that the road length is set to $\SI{56}{\meter}$, lane width to $\SI{13}{\meter}$ and the speed limit to $\SI{5.6}{\meter\per\second}$. Vehicles are modeled as point-masses and are required to drive within the lane. To avoid collision, a minimum required safety distance of $\SI{5.6}{\meter}$ between vehicles is enforced.
% \begin{figure}[t]
%   \centering
%   \includegraphics[width=1.0\linewidth]{Figures/three_vehicle.png}
%   \caption{Illustration of the evaluation scenario involving three vehicles on a highway.}
%   \label{fig:three_vehicle}
% \end{figure}

\textbf{Initial State Distribution.}
In order to evaluate the performance of each method, we consider a wide variety of initial conditions.
To focus on more challenging scenarios, i.e. those with conflicting objectives, we construct the set of initial conditions as follows.
First, we generate 10 base scenarios at which at least one vehicle cannot achieve all of their preferences perfectly.
We then sample 10 additional initial states from a uniform distribution centered around each base scenario.
We thus obtain a total 100 challenging scenarios.

\textbf{Evaluation Metrics.}
For each of these test problems, (\romannumeral 1) we evaluate methods based on the preferences \emph{at each priority level} for each player and (\romannumeral 2) we measure the $L_1$ distance between the trajectories found by each method.
To account for the existence of multiple equilibria, we solve the \ac{goop} 20 times, each time using a different initial guess.
We report the distance between the baseline trajectory and the \emph{closest} \ac{goop} trajectory. 
%Ideally, it would be best to compare the \ac{kkt} residuals \cref{eqn:goop-micp}. But this is not straightforward since \ac{goop} and the baselines do not share the same intrinsic dual variables. Instead, we resort to comparing the primals, which are common in both approaches. This provides a fair means of evaluating the performance of our method in computing \ac{goop} solutions

% \dongho{state we are checking two perspectives here: (1) reflecting ordered preferences (illustrating , and (2) distance to GNEP / trajectory-space-distance-plot (compare the primals, i.e. compare trajcetories) Ideally, it would be best to compare the KKT residuals but this is not readilly done since GOOP and baseline do not share the same duals.}
\subsection{Baseline: Explicitly Weighting Preferences}
When one does not have access to a solver capable of encoding preference hierarchies explicitly---the key feature of our proposed approach---one may instead attempt to encode the concept of ordered preferences via scalarized objective.
A natural scalarization scheme is a weighted sum of objectives per player---a technique that as been previously explored by \cite{veer2023receding} in non-game-theoretic motion planning.
We use a game-theoretic variant of this approach as a baseline.
Thus, for the baseline, each player solves a problem of the following form:
\begin{subequations}
\label{eqn:penalty-baseline}
\begin{align}
\min_{\bvar^i} \quad & \alpha_1 \cost{i}_1(\bvar^i, \bvar^{\neg i}) + \alpha_2 \cost{i}_2(\bvar^i, \bvar^{\neg i}) + \alpha_3 \cost{i}_3(\bvar^i, \bvar^{\neg i}) \label{eqn:penalty-objective}\\
\st \quad & \equality^i(\bvar^i, \bvar^{\neg i}) = 0, ~\inequality^i(\bvar^i, \bvar^{\neg i}) \geq 0, \label{eqn:penalty-equality-inequality} \\
    & \equality^s(\bvar) = 0, ~\inequality^s(\bvar) \geq 0. \label{eqn:penalty-shared-equality-inequality}
\end{align}
\end{subequations}
Here, $[\alpha_1, \alpha_2, \alpha_3] = [1, \alpha, \alpha^2]^\top$ in \cref{eqn:penalty-objective} is the vector of penalty weights assigned to each prioritized preference. 
To encode relative importance analogously to the hierarchical formulation in \cref{eqn:goop-K-level}, we choose $\alpha > 1$.

\textbf{Baseline variants.}
Observe that, for large penaty weights, the scalarized objective~\cref{eqn:penalty-objective} ensures a large separation of preferences at different hierarchy levels. % encodes a high relative importance of more important preferences over less important ones.
Hence, one may be tempted to choose $\alpha \gg 1$.
However, large penalty weights negatively affect the conditioning of the problem \cref{eqn:penalty-baseline}.
%Furthermore, it is not straighforward to determine  $\alpha$ the hierarchy will be (sufficiently) respected.
Since it is not straightforward to determine the lowest value of $\alpha$ that enforces the preference hierarchy, we instead consider several variants of the baseline with $\alpha \in \{1, 10, 20, 30, 40, 50\}$.
\subsection{Implementation Details}

We implement \cref{algorithm} and the aforementioned baseline in the Julia programming language.\footnote{Source code is available at \url{https://github.com/CLeARoboticsLab/ordered-preferences.}}
% \url{https://github.com/CLeARoboticsLab/ordered-preferences.}}
To ensure a fair comparison, we implement all methods using the same \ac{micp} solver, namely PATH~\cite{dirkse1995path}.

\textbf{Non-smooth objectives.}
Note that some of the objectives are not smooth, cf. \cref{eqn:runexp-reach-goal,eqn:runexp-speed-limit}, posing a challenge for numerical optimization.
However, since these objectives take the form $\cost{i}_{k}(\bvar^i_{k}, \bvar^{\neg i}_1) \coloneqq \max\big(0, -f^i_k(\bvar^i_{k}, \bvar^{\neg i}_1)\big)$, we can introduce a slack variable transformation to obtain a smooth problem, $\ie$, we can reformulate $\min_{\bvar^i_{k}} \max\big(0, -f^i_k(\bvar^i_{k}, \bvar^{\neg i}_1)\big)$ as:
\begin{subequations}
\label{eqn:smoothening-max-objective}
\begin{align}
    \quad \min_{\bvar^i_k, s_i^k} \quad & s^i_k \\
    \st \quad & s^i_k \geq -f^i_k(\bvar^i_{k}, \bvar^{\neg i}_1), \\
    \phantom{\st \quad} & s^i_k \geq 0. 
\end{align}
\end{subequations}

%\david{discuss how a hard cosntrained version would choke if the constraint is infeasible bt ours will not. remark with pointer to running example + result figure showing how we don't choke}

\subsection{Large-Scale Quantitative Results}
\label{sec:results-mc}
\cref{tab:comparison} shows the performance gap between our method (\acf{goop} \cref{eqn:goop-K-level} as implemented by \cref{algorithm}) and the baseline variants (game \cref{eqn:penalty-baseline}) with different penalty parameters.
Here, $\tilde{J}_k$ and $J_k$ denote the performance at preference level $k$ for the \emph{baseline} and \emph{our} method, respectively.

% \begin{table}[!t]
% \caption{Difference of preferences at each priority level\label{tab:comparison}}
% \centering
% \begin{tabular}{c c c}
% \hline
% \rule{0pt}{3ex}
%      & $\tilde{J}_3 - J_3$ & $\tilde{J}_2 - J_2$ \\ %[1.0ex]
% \hline
% R1 (Ambulance)  & $a \pm b$      & $a \pm b$      \\
% R2 (Passenger Car)   & $a \pm b$      & $a \pm b$      \\
% R3 (Passenger Car)   & $a \pm b$      & $a \pm b$      \\
% \hline
% \end{tabular}
% \end{table}

\begin{table}[h!]
\centering
\caption{Difference of preferences values at each priority level across different $\alpha$.}
\label{tab:comparison}
\begin{tabular}{@{}ccccc@{}}
\toprule
\textbf{Robot} & \textbf{$\alpha$} & $\tilde{J}_3 - J_3$ & $\tilde{J}_2 - J_2$ \\ \midrule
\multirow{4}{*}{R1 (Ambulance)} & 1  & 0.168 ± 0.004 & -1.76 ± 0.54 \\
                     & 10 & 0.179 ± 0.003 & -1.87 ± 0.54 \\
                     % & 20 & 0.058 ± 0.031 & -0.44 ± 0.59 \\
                    & 30 & 0.043 ± 0.044 & -0.22 ± 0.70 \\
                    % & 40 & 0.038 ± 0.042 & -0.04 ± 0.84 \\
                    & 50 & 0.046 ± 0.055 & -0.01 ± 1.03 \\ \midrule
\multirow{4}{*}{R2 (Passenger Car)} & 1  & 0.000 ± 0.001 & 0.00 ± 0.01 \\
                    & 10 & 0.000 ± 0.001 & 0.00 ± 0.01 \\
                    % & 20 & 0.000 ± 0.001 & 0.00 ± 0.01 \\
                    & 30 & 0.000 ± 0.001 & 0.00 ± 0.01 \\
                    % & 40 & 0.000 ± 0.001 & 0.00 ± 0.01 \\
                    & 50 & 0.002 ± 0.024 & 0.00 ± 0.01 \\ \midrule
\multirow{4}{*}{R3 (Passenger Car)} & 1  & 0.00 ± 0.00 & 0.00 ± 0.00 \\
                    & 10 & 0.00 ± 0.00 & 0.00 ± 0.00 \\
                    % & 20 & 0.00 ± 0.00 & 0.00 ± 0.00 \\
                    & 30 & 0.00 ± 0.00 & 0.00 ± 0.00 \\
                    % & 40 & 0.00 ± 0.00 & 0.00 ± 0.00 \\
                    & 50 & 0.00 ± 0.00 & 0.00 ± 0.00 \\ 
\bottomrule
\end{tabular}
\end{table}

\begin{figure}
  \centering
  \includegraphics[width=1.0\linewidth]{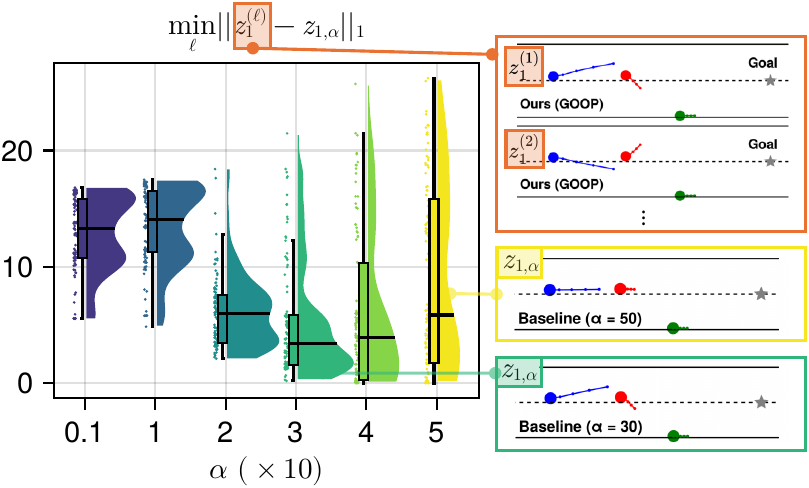}
  \caption{$L_1$-trajectory distance between GOOP solutions and baseline approximations at penalty strength~$\alpha$ ($z_{1,\alpha}$) for the 3-vehicle ambulance scenario. To facilitate a fair comparison when multiple solutions exist, we compute the distance as the \emph{minimum} difference between the baseline trajectory and \emph{all} available GOOP equilibria computed for the same initial condition. Note $z_1^{(\ell)}$ refers to the $\ell^\mathrm{th}$ GOOP trajectory for the given initial condition, where $\ell=1,2,\dots,20$.
  }
  \label{fig:trajectory-space-distance}
\end{figure}

%\textbf{Solver Reliability.}
Out of 100 test cases, \cref{algorithm} did not converge for six of the initial conditions at which the three vehicles were approximately collinear; we hypothesize that these instances correspond to boundaries between homotopy classes.
Therefore, the results below reflect only the remaining 94 test cases.
For reference, the baselines converged for all cases.

\textbf{Main Result 1: Preference Prioritization in GOOP.}
\cref{tab:comparison} shows the performance gap with respect to the multiple preference levels.
We see that the performance gap at the highest preference level (level ~3) is always positive (up to solver precision), indicating that our method finds solutions that perform better with respect to the highest priority preference.
Furthermore, \cref{tab:comparison} indicates that our method achieves this performance by ``backing down'' on lower priority preferences as indicated by the largely negative gap with respect to this metric.
In sum, these results support the claim that our method respects the order of preferences: \ac{goop} solutions relax less important preferences in favor of more important ones.

\textbf{Baseline performance.}
The baseline attenuates the performance gap as the penalty parameter $\alpha$ increases.
However, even with the largest penalty weight, $\ie$, $\alpha = 50$, the baseline fails to consistently match our method's performance and exhibits a high variance.
This effect can be attributed to poor numerical conditioning of the problem for large weights.

\textbf{Main Result 2: Distance between baseline and GOOP solutions}
\cref{fig:trajectory-space-distance} measures the $L_1$ distance between the baseline and \ac{goop} equilibrium trajectories for each test case. 
Although higher $\alpha$ values occasionally improve the baseline performance (as the lower end of the distributions approaches zero), for sufficiently high values of $\alpha$ the baseline exhibits poor numerical conditioning, resulting in a large variance in the solution quality, $\ie,$ the scalarized approximations do not always recover the \ac{goop} equilibria. 
This result shows the limitations of approximating \ac{goop} solutions via scalarization.

% In sum, this suggests that \ac{goop} framework is more appealing as it directly provides solutions aligned with ordered preferences without requiring tuning parameters like $\alpha$.}

% \textbf{Runtime statistics.}
% \dongho{are we adding this?} 
% Here are the runtime stat:...

%\textbf{Priority Leakage.}
%\lasse{TODO; add here or further down below: to further support the claim that our method strictly respects the preference hiearchy, we should have a figure that supports our key claim (as indicated in our old title): in constrast to the baseline, our method never backs down unless it has to. One way to visualize the number of levels that each method enforces "exactly" (within a certain tolerance of optimality)}
%Indeed, even with high $\alpha$, the baseline algorithm is sensitive to lower-priority preferences, as shown in \cref{fig:highway-expmt-mc-level2} where large negative differences in the players' aggregate performance at level $2$ indicate that the baseline sacrifices performance at level $3$ in order to improve performance at level $2$.

\begin{figure}
  \centering
  \includegraphics[width=0.75\linewidth]{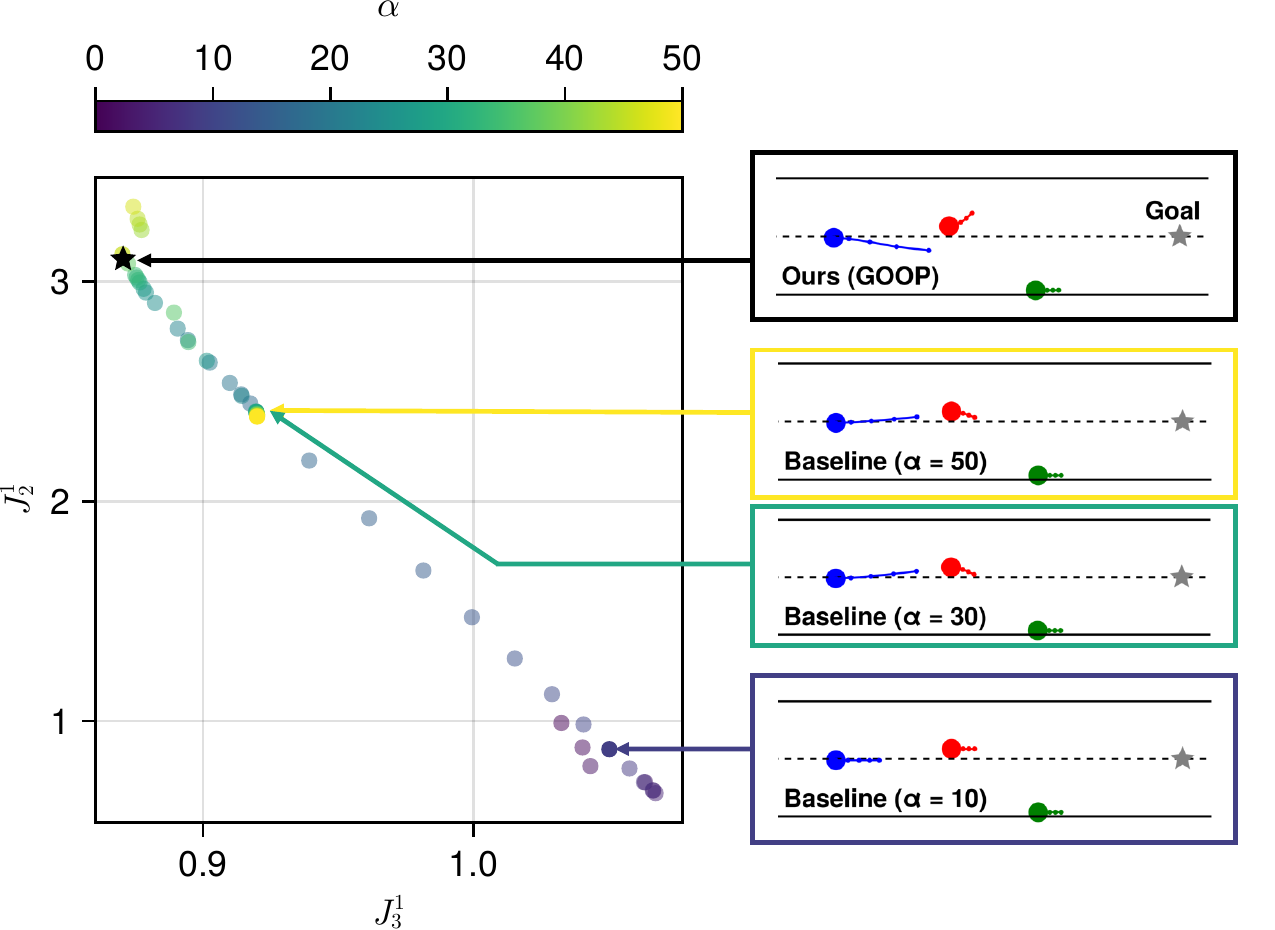}
  \caption{Comparison of vehicle 1's highest (get to goal) and second highest (obey speed limit) preference values for \ac{goop} and baseline for different values of $\alpha$.
   Increasing $\alpha$ initially improve the trajectory for R1. However, the performance improvement is not monotonic since at $\alpha = 30$ and $\alpha = 50$, the baseline yields a degraded trajectory for R1, $\ie$ farther away from the goal position.}
  \label{fig:pareto-front}
\end{figure}

%Our results also indicate that the baseline does not always monotonically approach the performance of \cref{algorithm} as $\alpha \to \infty$. 

\subsection{Detailed Analysis for a Fixed Scenario}
To provide additional intuition beyond the large-scale evaluation in \cref{sec:results-mc}, next, we assess a \emph{fixed} scenario in greater detail.
\cref{fig:pareto-front} visualizes the solutions identified by both \cref{algorithm} and the baseline for a single initial state and a dense sweep over the penalty weight, $\ie$, $\alpha \in \{1, 2, \dots, 49,  50\}$. 
In \cref{fig:pareto-front}, we plot player 1's preferences at level~3 $(J_3^1)$ over their preferences at level~2 $(J_2^1)$.
To illustrate how the solutions in \cref{fig:pareto-front} correspond to open-loop trajectories, we link selected points to their respective trajectories on the right side.

\textbf{Quantitative Results.}
Our \ac{goop} solution, marked by a star at the top left, outperforms all baselines, achieving the lowest value of preference at the most important level. 
All baseline solutions are located to the right of the \ac{goop} solution, indicating that the baselines do not consistently match \ac{goop} in optimizing the highest priority preference. 
In line with the large-scale evaluation in \cref{sec:results-mc}, we observe that larger weights do not consistently improve performance.
In fact, $\player{1}$'s trajectory worsens for $\alpha = 30$ and $\alpha = 50$. 
%This performance degradation under higher $\alpha$ highlights how increasing the penalty weight on certain preferences does not yield uniformly better outcomes. 

\textbf{Qualitative Results.}
Recall that for $\player{1}$, reaching its goal has the highest priority.
By accurately encoding this prioritization, our method finds a solution that brings $\player{1}$ closer to the goal at the final time step than all baseline variants.
For $\player{2}$ and $\player{3}$, all methods achieve comparable performance with respect to all prioritized preferences.
In summary, these results further support the claim that the equilibrium solutions computed by \cref{algorithm} reflect players' hierarchical preferences.

\subsection{An Intersection Scenario}
We present an intersection scenario involving two vehicles, each with \emph{four} levels of preferences.
% Specifically, vehicle $1$ (ambulance) prioritizes goal reaching over driving within its designated lane, which is, in turn, prioritized over obeying the speed limit (of $\SI{10}{\meter\per\second}$), and minimizing control effort.
% Vehicle $2$ (passenger car) prioritizes driving within its designated lane over obeying the speed limit, which is, in turn, prioritized over reaching the goal, and minimizing control effort.
In accordance with the order of preferences outlined in \cref{fig:goop-intersection}, vehicle $1$ accelerates beyond the speed limit (of $\SI{10}{\meter\per\second}$). 
In contrast, vehicle $2$ maintains its speed but sacrifices reaching the goal.
Both vehicles achieve the top preference by sacrificing their less important preferences, showing that our \ac{goop} framework accurately captures the hierarchy of preferences in settings with deeply nested objectives.
\label{sec:results}

\section{Conclusion and Future Work} \label{sec:conclusion}
In this paper, we proposed the \acf{goop}, a multi-agent, noncooperative game framework where each player optimizes over their own hierarchy of preferences. 
We recursively derived first-order optimality conditions for each player's optimization problem, which introduces complementarity constraints. 
We proposed a relaxation-based algorithm for solving the $N$-player KKT system for approximate (local) \ac{gne} solutions. 
Our experiments show that our algorithm outperforms penalty-based baselines while accurately reflecting each individual's order of preferences by relaxing lower-priority preferences when needed.
%Our results demonstrate that \ac{goop} solutions automatically relax less important preferences when more-important ones are at ric

%Future research may develop more tailored relaxation techniques for solving \acp{goop} while maintaining numerical stability. 
Future work may focus on tailored numerical solvers that avoid the need for iteratively solving relaxed \ac{micp} subproblems. Our formulation's problem size grows exponentially in the number of hierarchy levels. While real-time performance is not the immediate goal of this work, future work should address this limitation, in which case our work can serve as a reference solution technique.
Future work may also explore amortized optimization via neural network policies that approximate equilibrium solutions, potentially enabling larger-scale deployments with many players and deeper preference hierarchies.
Finally, extending \ac{goop} to incorporate feedback mechanisms for dynamic information presents an exciting opportunity for applications such as autonomous mobile agents.

\bibliographystyle{IEEEtran}
{\footnotesize\bibliography{6_ref}}

% \newpage

 

% \begin{IEEEbiography}[{\includegraphics[width=1in,height=1.25in,clip,keepaspectratio]{Figures/highway.png}}]{Michael Shell}
% \end{IEEEbiography}



\vfill

\end{document}